\documentclass[]{spie}  

\usepackage{amsmath,amsfonts,amssymb}
\usepackage{graphicx}
\usepackage[colorlinks=true, allcolors=blue]{hyperref}

\title{Design of an IR Imaging Channel for the Keck Observatory SCALES Instrument}

\author[a]{Ravinder K. Banyal}
\author[a]{Amirul Hasan}
\author[b]{Reni Kupke}
\author[a]{Hari Mohan Varshney}
\author[a]{Ajin Prakash}
\author[a]{T. Sivarani}
\author[b]{Andy J. Skemer}
\author[b]{Nick MacDonald}
\author[c]{Steph Sallum}
\author[b]{Will Deich}
\author[d]{Michael P. Fitzgerald}
\author[a]{K. V. Govinda}
\author[b]{Chris Ratliff}
\author[a]{Ramya Sethuram}
\author[b]{Deno Stelter}
\author[f]{Arun Surya}
\author[d]{Eric Wang}
\affil[a]{Indian Institute of Astrophysics, INDIA}
\affil[b]{University of California, Santa Cruz, USA}
\affil[c]{University of California, Irvine, USA}
\affil[d]{University of California, Los Angeles, USA}
\affil[e]{W. M. Keck Observatory, USA}
\affil[f]{Tata Institute of Fundamental Research, INDIA }

\authorinfo{Further author information: (Send correspondence to RKB)\\RKB: E-mail: banyal@iiap.res.in, 
Telephone: +918022541272\\  A.J.K.: E-mail: askemer@ucsc.edu, Telephone: +1 831-459-5753}

\begin{document} 
\maketitle

\begin{abstract}
A next-generation instrument named, Slicer Combined with Array of Lenslets for Exoplanet Spectroscopy (SCALES), is being planned for the W. M. Keck Observatory. SCALES will have an integral field spectrograph (IFS) and a diffraction-limited imaging channel to discover and spectrally characterize the directly imaged exoplanets. Operating at thermal infrared wavelengths (1-5~$\mu$m, and a goal of 0.6-5~$\mu$m), the imaging channel of the SCALES is designed to cover a $12''\times12''$ field of view with low distortions and high throughput. Apart from expanding the mid-infrared science cases and providing a potential upgrade/alternative for the NIRC2, the H2RG detector of the imaging channel can take high-resolution images of the pupil to aid the alignment process. Further, the imaging camera would also assist in small field acquisition for the IFS arm. In this work, we present the optomechanical design of the imager and evaluate its capabilities and performances.
\end{abstract}

\keywords{Exoplanet atmospheres, high-contrast imager, integral field spectrograph, Keck instruments}

\section{INTRODUCTION}
\label{sec:intro}  
The search for extrasolar planets is largely driven by humanity's quest to find evidence of life outside the Earth. To this date, over five thousand extrasolar planets have been discovered using various methods. Each search method probes different regions of  planet-star parameter space, encompassing a vast range of planetary mass, radius, orbital distance and age [\citenum{fis15}]. Besides discovering new planets, now the focus is to study the planetary atmospheres and to look for the possible signatures of biological molecules indicating the presence of life [\citenum{dan18, wal18}].  Early discoveries of exoplanets mostly came from radial velocity and transit surveys.  With the arrival of new technology, it became possible to directly image and characterize the young and warm planets in the outer orbits of the stars [\citenum{mar06}]. A small angular separation ($\leq0.1^{\prime\prime}$ ) and a huge luminosity difference ($\leq 10^{-6}$) between the planet and the star pose a major difficulty for the direct imaging. This challenge is addressed by high-contrast instruments relying on the ability of a coronagraph to curtail the starlight and the adaptive optics (AO) technology to push the performance of telescopes to the near diffraction-limit. The recently lunched James Webb Space Telescope’s NIRCam and MIRI instruments have coronagraphic capabilities  at near- and mid-infrared wavelengths [\citenum{boc05}]. Several high-contrast instruments, such as, GPI/Gemini [\citenum{gra14}], SPHERE/VLT [\citenum{beu08}], NIRC2/Keck [\citenum{maw18}], and  CHARIS/Subaru [\citenum{lim13}]  are currently operating on large ground telescopes. These instruments provide spatially resolved images and spectra of exoplanets at visible and infrared wavelengths.

SCALES is a facility class instrument planned for the W. M. Keck Observatory to do integral field spectroscopy and high-contrast imaging of astronomical objects over a narrow field. The instrument will operate at thermal infrared ($2-5\mu$m) band where detecting young and self-luminous exoplanets is easier due to  lower brightness contrast. Among the existing exoplanet instruments,  SCALES has an overall advantage in terms of spectral resolution, imaging contrast, and wavelength coverage. Although  the primary goal of the SCALES is to discover and characterize exoplanets, its  research capabilities will extend to solar system objects, active galactic nuclei, protoplanetary discs, and star-forming regions.   For recent updates on SCALES, reader can  refer to a review paper (\#~12184-18) by A. Skemer published in these proceedings [\citenum{ske22}]. 

\section{Instrument Overview}

\begin{figure} [ht]
   \begin{center}
   \begin{tabular}{c} 
   \includegraphics[height=7cm]{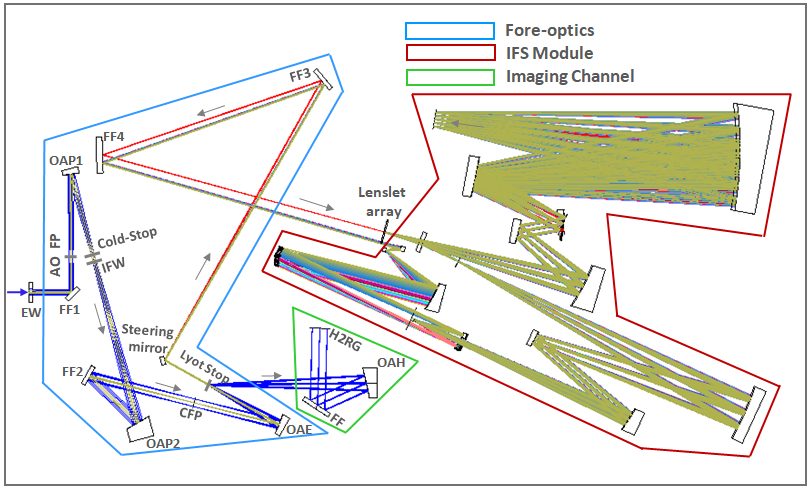}
   \end{tabular}
   \end{center}
   \caption[example] 
   { \label{fig1} 
The optical layout of SCALES with three subsystems.  The foreoptics (blue-box) reshapes the telescope beam to provide a set of  image and pupil planes for the placement of cold-stop, imager filter wheel, coronograph occulters, Lyot stop and lenslet array. The  IFS module (red-box), receives the light from lenslet array to do low- and mid-resolution spectroscopy. A mirror inserted in the Lyot wheel deflects the light into  imaging channel (green-box). The components labeled in the figure are EW: Entrance Window; FF\#: Flat Fold; AO FP: Adaptive Optics Focal Plane; OAP\#: Off-axis-parabola; IFW: Imager Filter Wheel; CFP: Coronagraphic Focal Plane; OAE: Off-Axis-Ellipse; H2RG: HAWAII-2RG detector. }
   \end{figure}

The optical design of the SCALES is shown in Fig.~\ref{fig1}. Functionally, the optical layout of the instrument is divided into three parts: 1) Fore-optics; 2) IFS module; and 3) Imaging Channel. The foreoptics (outlined in blue) consists of a number of flat fold mirrors and off-axis conics  which create  a set of image and pupil planes required to place key coronagraphic components and relay the beam downstream to the IFS and imaging channels.  The foreoptics is common for both spectrograph and imager.

The AO corrected beam from Keck enters SCALES from the entrance window on the left, as shown in Fig.~\ref{fig1}. The entrance window also provides a vacuum interface for optical beam propagating into the SCALES' cryostat. The first AO focal plane is located on the instrument bench following the reflection from FF1. The OAP1 creates a high quality pupil image (image of M1) where  a rotating cold-stop is located to block the thermal noise from the observatory and telescope structure. The imager filter wheel which contains a set of bandpass filters resides in the collimated space adjacent to the cold stop. The OAP2 creates the coronagraphic focal plane where a linear slide mechanism holding several masks, and a pupil imaging lens which is used for aligning the cold stop to the telescope primary. The aspheric OAE provide a well-formed pupil image for Lyot stop for pupil apodization. The Lyot wheel is additionally  equipped with a fold mirror to feed the imaging channel (outlined in green). The OAE, further magnifies the beam while reflecting from a series of flat mirrors (steering mirror, FF3 and FF4) to create a downstream image plane (f/350 beam) that is sampled by the lenslet array at the front end of the IFS channel (outlined in red).

The f/8 output beam from the lenslet array is relayed to the spectrograph optics (not labeled in the Figure) designed to do one-to-one imaging. The IFS channel has two selectable modes, namely,  the low-resolution IFS and the mid-resolution IFS. In low-resolution mode, the light from $110 \times 110$ lenslet array (FoV= $2.2''\times 2.2''$) goes directly to the spectrograph collimator and dispersed by a prism.  Tilted and interleaved spectra for each lenslet micropupil are imaged by the spectrograph camera on $2\mathrm{K}\times2\mathrm{K}$ H2RG detector. The mid-resolution mode utilizes a unique design concept to re-formats light from $18\times 18$ grid of micropupils (FoV = $0.36''\times0.36''$) to form three pseudo slits which allow dispersion of the 306 spaxels  across the full length of the detector.  The tip-tilt steering mirror is used for precise centring and holding the exoplanet image either on the low-resolution or the mid-resolution part of the lenslet array. The fore-optics upstream of the Lyot wheel is oversized and designed to provide a diffraction-limited unvigneted field of view ( $>20''\times20''$)  for the imager. Reflective optics for all subsystems of SCALES, including the imager, will be made from the diamond-turned Al metal (RSA-Al-6061 or equivalent) due to its advantage of the more refined grain structure for optical surfaces. Likewise, the mounts and optical benches will be machined from Al 6061. The entire instrument is to be assembled on an optical bench, vacuumed and cooled to cryogenic temperatures (77~K).  Many system level details of the IFS baseline  instrument are covered in previous publications  [\citenum{den20, gon21, den21, li21}]. Full description of the optical design of the SCALES  can be found in  R. Kupke's paper (\#~12184-159) appeared in the current proceedings [\citenum{kup22}]. In this paper, we present the imaging channel design which consists  of four subsystems: Imager Optics, Imager Filter Wheel, Pupil Mask Rotator and H2RG Detector System.  

   \begin{figure} [ht]
   \begin{center}
   \begin{tabular}{c} 
   \includegraphics[height=6.5cm]{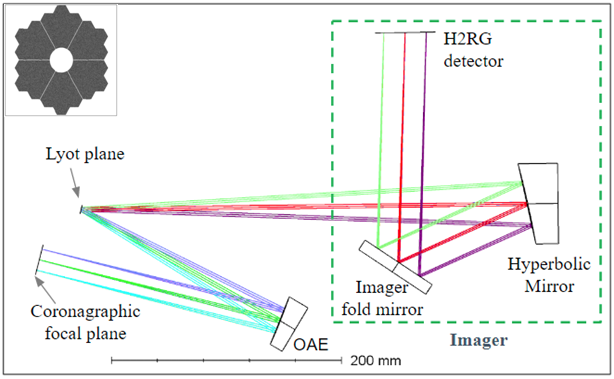} \\
   \end{tabular}
   \end{center}
   \caption[example] 
   { \label{fig2} 
The Zemax model of the imaging channel. For taking science images, a flat mirror in the Lyot plane reflects the light into the imager optics consisting of a hyperbolic mirror and  a fold mirror. Pupil imaging mode is realized using a planoconvex  lens placed at the coronagraphic focal plane and a spherical mirror at the Lyot plane. Pupil image shown in the inset is obtained using Zemax's geometric image simulation feature.  
}
   \end{figure} 

\section{Rationale for Imaging Channel} 
During the past two decades, the near-infrared camera, NIRC2, on Keck has made important contributions to high-contrast imaging, enabling new discoveries in exoplanet science, galactic astronomy, stars and solar system objects. The SCALES imager channel is designed to match (and in some cases exceed) the performance of  NIRC2, which is now aging and needs to be replaced in near future. The science capabilities of NIRC2 are used as the primary driver to arrive at the top-level  design requirement for the SCALES' imager. Apart from expanding the mid-infrared science cases and providing a potential alternative for the NIRC2, imaging channel will take high-resolution images of the pupil to aid the alignment process. Further, the H2RG camera of the imager would also assist in small field acquisition for the IFS arm. The imager optics is, therefore, designed to provide functionality: a) for taking science images of astrophysical objects and b) for imaging the pupil on H2RG detector. 

\section {Imager Optics Design}
The main optical requirements for the imaging channel are listed in Table~\ref{tab1}. Imager functioning across the 1-5 micron bandpass, provides a diffraction-limited image quality  at shortest operating wavelength of 1~$\mu$m. The FoV, pixel scale, and RMS WFE  are designed to be better than the  narrow field ($10''\times10''$) camera of the NIRC2. The requirement for distortion and telecentricity mostly come from astrometric precision needed for galactic centre observations. It is important that distortions are low-order so that they are easy to calibrate and correct.

\begin{table}[ht]
\caption{Optical requirements for the imaging channel.} 
\label{tab1}
\begin{center}       
\begin{tabular}{|l|l|l|} 
\hline
\rule[-1ex]{0pt}{3.5ex}  Parameter & Requirement (Goal) & Achieved \\
\hline
\rule[-1ex]{0pt}{3.5ex}  Image quality & Nyquist sampled @1$\mu$m (0.6 $\mu$m) & yes  \\
\hline
\rule[-1ex]{0pt}{3.5ex} Wavelength range &  1-5 $\mu$m (0.6-1 $\mu$m) & yes\\
\hline
\rule[-1ex]{0pt}{3.5ex}  Field of View (FoV)      & $>10''\times10''$ ($20''\times20''$) & $12.3''\times12.3''$  \\
\hline
\rule[-1ex]{0pt}{3.5ex}  Plate scale    & $< 0.01$ arcsec/pix (0.006 arcsec/pix) & 0.006 arcsec/pix \\
\hline 
\rule[-1ex]{0pt}{3.5ex}  RMS WFE  & $< 190$ nm  (60-70 nm) & $<65$ nm \\
\hline 
\rule[-1ex]{0pt}{3.5ex}  Atelecentricity  & $< 0.057$ deg (0.038 deg) & 0.04 deg \\
\hline 
\rule[-1ex]{0pt}{3.5ex}  Distortion  & $\leq 0.045$ arcsec, low-order (0.03 arcsec) & $0.045$ arcsec, low-order  \\
\hline
\rule[-1ex]{0pt}{3.5ex}  Filters  & YJHKL + NIRC2 filters & yes \\
\hline
\rule[-1ex]{0pt}{3.5ex} Pupil image size & $> 300$ pixel & 455 pixel \\
\hline
\rule[-1ex]{0pt}{3.5ex}  Pupil alignment precision  & $\leq 1\%$ & -- \\
\hline
\end{tabular}
\end{center}
\end{table}

The Zemax model of the imager is shown in Fig.~\ref{fig2}. The telescope beam relayed by SCALES' forward optics is intercepted by a plane mirror mounted at $15^\circ$ angle in the Lyot wheel which deflects the light into the imaging channel that has two optical elements: an off-axis hyperbola (OAH) and a flat-fold mirror. These two elements form a final image on the H2RG detector, which is tilted by $2.58^\circ$ to the incoming beam,  ensuring lowest wave-front error and a uniform psf over the field of view. The optical design of the imager does not impact the baseline spectrograph instrument in any way.  Specifications of the imaging channel optics are provided in Table~\ref{tab3}. 

\begin{table}[ht]
\caption{Pupil specifications for SCALES at different locations.} 
\label{tab2}
\begin{center}       
\begin{tabular}{|l|l|l|l|} 
\hline
\multicolumn{1}{|c|}{Pupil} & \multicolumn{3}{|c|}{Obscuration size (dia)} \\
\cline{2-4}
\rule[-1ex]{0pt}{3.5ex} location & M1 (mm)  & M2 (mm) & Spider width  (mm) \\
\hline

\rule[-1ex]{0pt}{3.5ex} Entrance  & 10,000 & 2,480 & 25  \\
\hline
\rule[-1ex]{0pt}{3.5ex} Cold stop & 14.96 & 3.39 & 0.034 \\
\hline
\rule[-1ex]{0pt}{3.5ex} Lyot stop & 7.27 & 1.65 & 0.017  \\
\hline
\rule[-1ex]{0pt}{3.5ex} Detector  & 8.00  & 1.81 & 0.018  \\
\hline
\end{tabular}
\end{center}
\end{table}

\subsection{Pupil Imaging}
Table~\ref{tab2} provides the Keck II pupil dimensions at various locations inside the SCALES [\citenum{li21}]. For alignment purpose, the pupil image at the cold stop plane has to be re-imaged on the detector.  We first explored a signal lens design which required a separate mechanism to deploy a lens before the coronagraphic focal plane. However, due to space constraint, the single lens option was discarded and a two elements design was chosen to make use of the existing deployment mechanism at coronagraphic and Lyot planes (see e.g., Fig.~\ref{fig2}). The pupil imaging optics consists of a plano-convex lens mounted  in the coronagraphic slide and a spherical mirror in the Lyot wheel. The pupil image size on the detector is 8~mm ($\sim455$~pixels). There is no stringent requirement of the image quality but it is good enough to achieve $\sim 1\%$ alignment precision between the mask and the pupil.

   \begin{figure} [ht]
   \begin{center}
   \begin{tabular}{c} 
   \includegraphics[height=6.6cm]{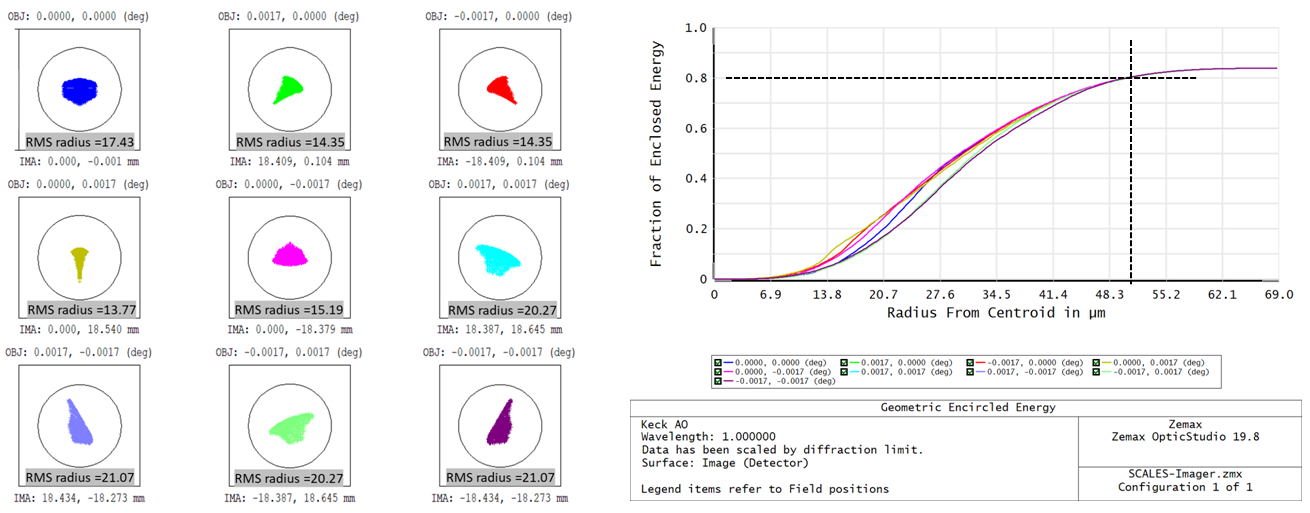}
   \end{tabular}
   \end{center}
   \caption
   { \label{fig3} 
Left panel: Diffraction-limited spots diagram for the imaging channel design. Black circles representing Airy disks of dia= 137.63~$\mu$m, calculated for $\lambda=1\mu$m. Different colors represent different field points. Right panel: Encircled energy distribution.}
   \end{figure}

\subsection{Optical Performance}
\begin{figure} [ht]
   \begin{center}
   \begin{tabular}{c} 
   \includegraphics[height=5.5cm]{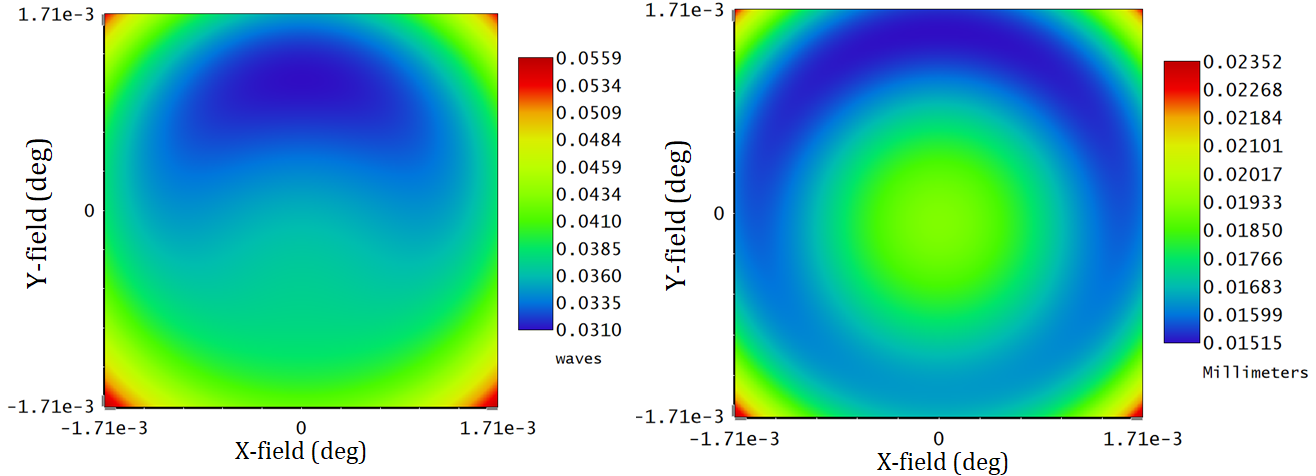}
   \end{tabular}
   \end{center}
   \caption
   { \label{fig4} 
Left panel: Distribution of RMS wavefront error at $\lambda=1\mu$m.  Right panel: Distribution of RMS spot radius.}
   \end{figure}

The optical design of the imager meets the technical requirements stated in Table~\ref{tab1}. Figure~\ref{fig3} (left panel) shows the diffraction-limited spot diagram for the various field positions.  The right panel in Fig.~\ref{fig3} shows the radial distribution of energy where 80\% encircled energy is confined within a $51 \mu$m radius. The rms wavefront error and rms spot radius maps are shown in Fig.~\ref{fig4}. The maximum rms wavefront error at the corners is  $55.9$~nm while the difference across the field remains below $30$~nm. The rms spot radius over the field varies from minimum 15.15~$\mu$m and to maximum 23.52~$\mu$m.

\begin{figure} [ht]
   \begin{center}
   \begin{tabular}{c} 
   \includegraphics[height=6cm]{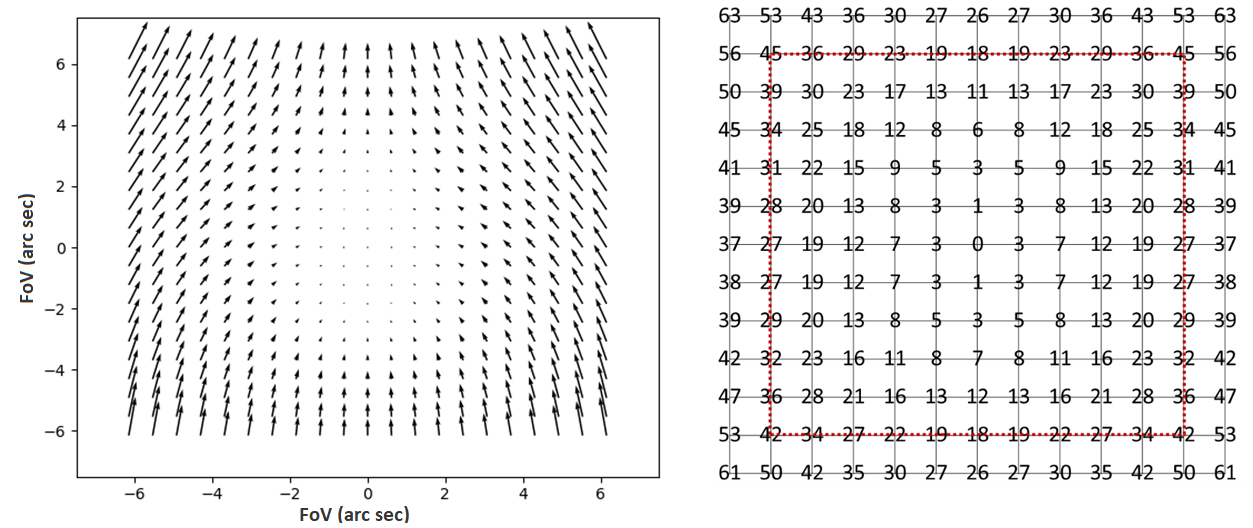}
   \end{tabular}
   \end{center}
   \caption[example] 
   { \label{fig5} 
Left panel: Distortion map of the imager ($20\times$ magnification).  Right panel: Numerical representation of distortions in milli arcsec. The red dotted square on the right corresponds to $10''\times10''$ field.}
   \end{figure}

The distortion maps covering $12.3''\times12.3''$ field (left panel)  and the corresponding numerical values mapped over 1~arcsec grid are shown in the right panel of Fig.~\ref{fig5}. The maximum distortion over $10''\times10''$ FoV is $0.045$~arcsec. Clearly, the field distortions are all low-order, which are easy to calibrate and correct. The maximum reported distortion for NIRC2 is 0.04~arcsec, but the spatial distribution is complex [\citenum{yel10}]. The performance of the pupil imaging system is not exactly diffraction-limited, though the 80\% of energy is confined within the 228 microns (13 pixels). The requirement is relaxed as we only need to resolve the outer segmented boundaries of the pupil to align it with a cold stop. The atelecentricity achieved for the imager design is 0.04 deg against the 0.057 deg requirement.

\begin{table}[ht]
\caption{Prescription of optical components used in the imager. }
\label{tab3}
\begin{center}       
\begin{tabular}{|l|l|l|l|l|} 
\hline
\rule[-1ex]{0pt}{6ex} S.No & Component & Focal length & Aperture & Conic \\
\hline
\rule[-1ex]{0pt}{3.5ex} 1. & Pupil imaging lens & 113.8 mm & 2 mm & 0 \\
\rule[-1ex]{0pt}{3.5ex}  & (plano-convex) &  & & 0 \\
\hline
\rule[-1ex]{0pt}{3.5ex} 2. & Lyot mirror (spherical) & 350 mm & 11 mm & -- \\
\hline
\rule[-1ex]{0pt}{3.5ex} 3. & Lyot mirror (flat) & $\infty$ & 11 mm & -- \\
\hline
\rule[-1ex]{0pt}{3.5ex} 4. & Fold mirror & $\infty$ & 70 mm & --   \\
\hline
\rule[-1ex]{0pt}{3.5ex} 5. & OAH & 369.4 mm & 70 mm & -1.096 \\
\hline
\end{tabular}
\end{center}
\end{table}

\section{Cold-stop Rotation Mechanism}
\begin{figure} [ht]
   \begin{center}
   \begin{tabular}{c} 
   \includegraphics[height=8.0cm]{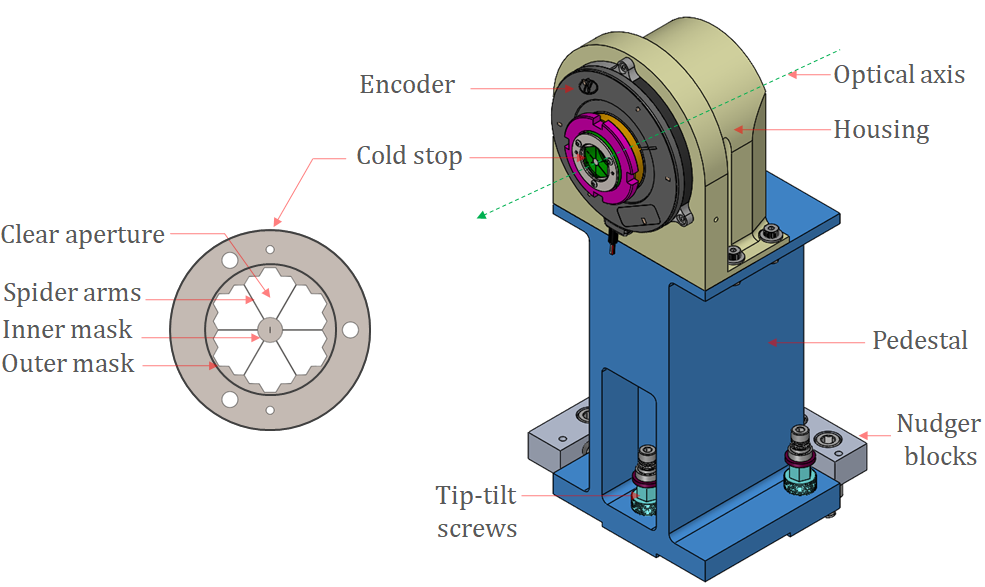}
   \end{tabular}
   \end{center}
   \caption
   { \label{fig6} 
A CAD model of the cold stop rotating mechanism. Left panel: A circular frame holding the  cold stop. Three large holes which are 120 deg apart are meant to bolt the cold stop frame to the rotating shaft. Two smaller holes are meant for locating the frame on the shaft. Right panel: A complete assembly with pedestal support.}
   \end{figure}

A cold stop matching the dimension and shape of the pupil is required to block the unwanted thermal emission from the telescope structure (spider arms, M2 obscuration and support) seen in the pupil image.  Depending on the science needs, the Keck AO system can either provide a \textit{pupil tracking mode} (fixed pupil with a rotating field), or a \textit{field tracking mode} (fixed field with a rotating pupil). The former requires a static cold stop well aligned with pupil, while in the later case, the cold stop  must rotate synchronously with the pupil with a varying speed that depends on the altitude of the telescope. The CAD model of the cold stop rotator designed for the SCALES is shown in Fig.~\ref{fig6}. The dimensions of secondary obscuration, the spider arms and the outer mask are slightly oversized to account for the the Keck II pupil nutation which is approximately 1\% of the diameter of the primary mirror [\citenum{li21}]. For the finalized  cold-stop and Lyot stop design, readers can also refer to Jialin Li's paper in these proceedings [\citenum{li22}]. In order to reduce unwanted thermal noise upfront in the optical chain of the SCALES, the cold stop subsystem would be positioned at the 1st pupil image (see Fig.~\ref{fig1}) formed by OPA1.

\begin{figure} [ht]
   \begin{center}
   \begin{tabular}{c} 
   \includegraphics[height=8cm]{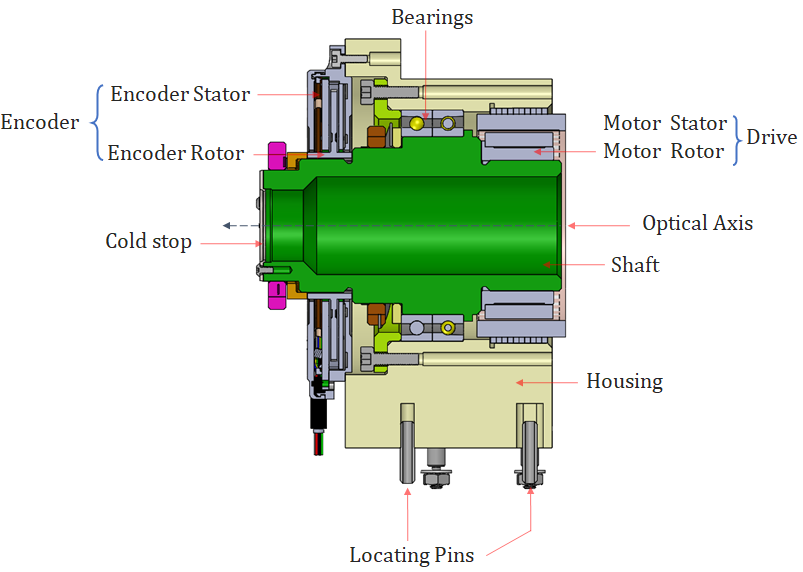}
   \end{tabular}
   \end{center}
   \caption
   { \label{fig7} 
A crossectional view of the cold stop rotator.}
   \end{figure} 

We plan to use a cold mask rotation concept similar to the one proposed  for HARMONI, which is an integral field spectrograph for the E-ELT [\citenum{gig16}]. A cutout view of the housing assembly of the rotator is shown in Fig.~\ref{fig7}. The mechanism consists of a shaft coupled to the housing with the help of a set of angular contact bearings arranged back-to-back. A brushless DC motor (Applimotion Celeram Model: UTS-53-A-20-AN-000) is attached to one end of the shaft and the cold stop frame to the other. This axial mounting scheme  requires no additional geared transmission which would have complicated the design. Additionally, the direct drive motor eliminates any backlash that could adversely affect the functionality of the instrument. Furthermore, the motor uses Samarium Cobalt (SmCo) magnets that outperforms Neodymium magnets at cryogenic temperatures. The motor is commercially  available off-the-shelf, though it is not certified by the manufacturer for usage in cryogenic environment. However, HARMONI/E-ELT team has done laboratory tests on a similar motor (different model), qualifying its operation at cryogenic temperatures without any functional issue [\citenum{gig16}].

An absolute encoder (model: Netzer DS-70) with a 19 bit resolution is chosen to provide accurate position and speed feedback for the rotator. This is a capacitive encoder with a hollow shaft that easily integrates axially with the main shaft. The encoder has been tested successfully down to 25K temperature and the errors generated in measurements due to temperature variations are repetitive which could be corrected by proper calibrations [\citenum{gig16}]. The housing is made of  stainless steel (SS 304) and circular slots are to be  milled with a single setting to achieve collinearity of the axis. The housing is attached to  aluminium  (Al6061-T6) pedestal which is supported on the bench. The housing is located on the pedestal with two locating pins that maintains it's position at all temperatures. Four springs coupled with bolts press the lower surface of the housing against the top surface of the pedestal unit thus maintaining the height of the mask axis from the bench surface. The base of the pedestal houses three screws to aid tip-tilt adjustments during alignment and assembly. A set of nudger blocks with set screws provide controlled linear movements in the plane of the bench surface. The bench acts as a heat sink and the pedestal also creates a continuous thermal path between the housing and the bench. The main heat generating element is the stator that has a direct conduction path to the housing which is thermally strapped to the bench to remove excessive heating. Since the instrument will go through cool down and vacuum cycles, it is designed to ensure all the holes are provided with air escape path to avoid air entrapment.    

\section{Imager Filter Wheel Design}
Residing next to the first pupil plane of the foreoptics,  a double stacked filter  wheel mechanism hosts 2 open slots and 16 filters matched to the various imaging bandpass of the imager. The maximum permissible distance from the pupil plane to maintain the diverging beam's footprint on the filter within $< 90\%$ clear aperture is 85~mm.

\begin{figure} [ht]
\begin{center}
\begin{tabular}{c}  
\includegraphics[height=8.5cm]{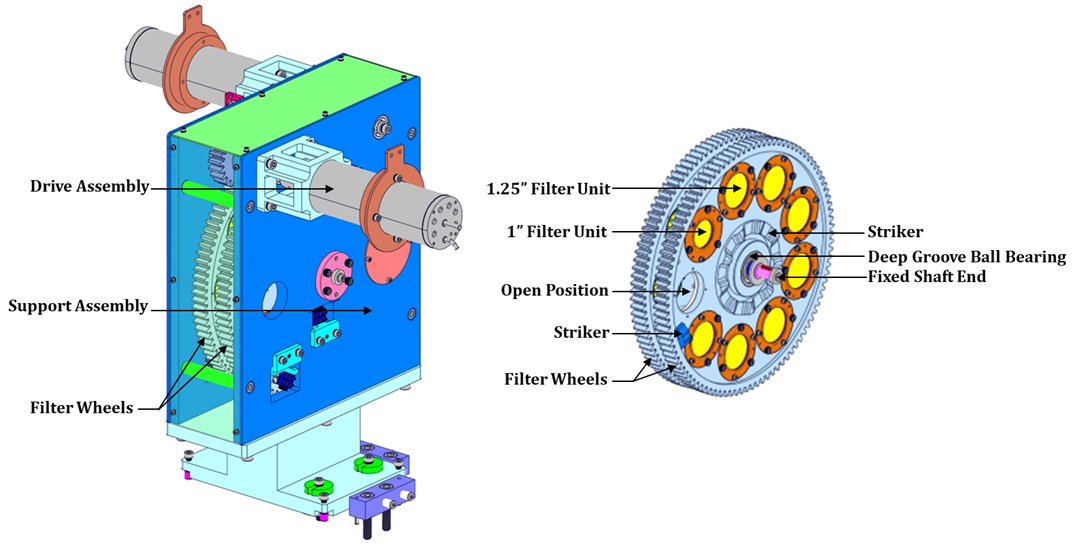}
\end{tabular}
\end{center}
\caption[example] 
   {\label{fig8} 
Left panel: A CAD model of the double stacked filter wheel. Right panel: Filter wheel assembly seen from inside.}
   \end{figure}

\begin{figure} [ht]
\begin{center}
\begin{tabular}{c}  
\includegraphics[height=5.0cm]{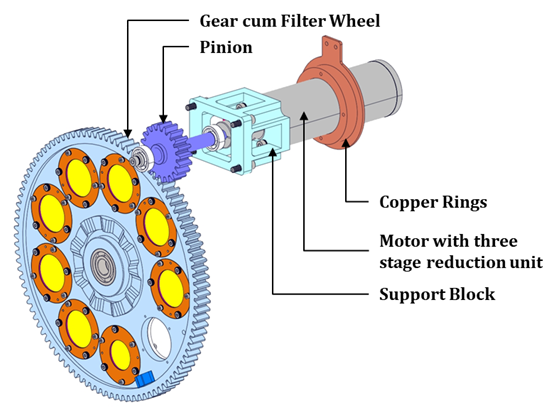}
\includegraphics[height=8.0cm]{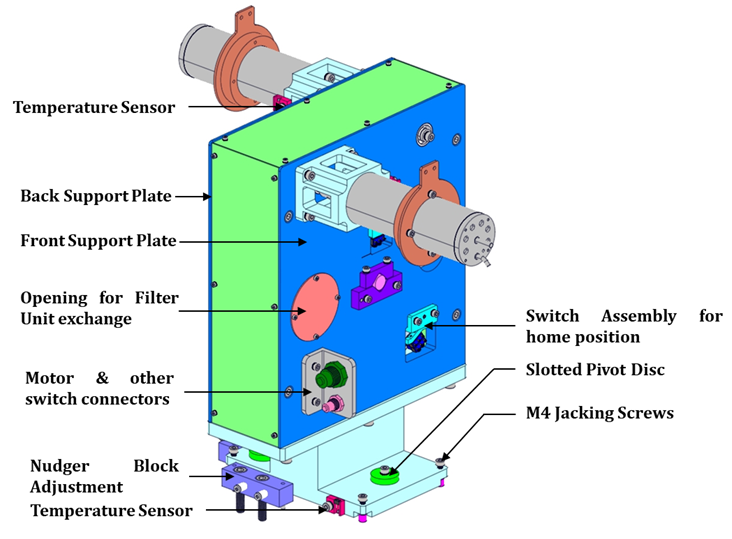}
\end{tabular}
\end{center}
\caption[example] 
   {\label{fig9} 
Left panel: Drive mechanism for the filter wheel. Right panel: Support assembly with electronic and mechanical parts. }
   \end{figure}

The mechanical design of the IFW  shown in Fig.~\ref{fig8} (left panel) has a support assembly holding two filter wheels and a drive mechanism for each. One end of the filter wheel shaft is fixed, and another has the freedom to slide along the shaft axis to compensate for the contraction and  expansion due to temperature differences. Each filter wheel seen in Fig.~\ref{fig8} (right panel) has 8 filter slots where 1-inch (standard) and 1.25-inch (used in NIRC2)  filter  can be mounted using shoulder screws guided by dowel pins.  Both filter wheels have separate drive mechanism, each consisting of a simple gear train with a three-stage inbuilt reduction unit stepper motor (pn VS43 200.2.5F). The simple gear train has a gear ratio of 5 and 70 for an inbuilt three-stage unit. The total gear ratio of the drive is 350.  The motor shaft is connected to the pinion shaft with a coupling to rotate the pinion, meshing with the filter wheel gear (see left panel of Fig.~\ref{fig9}). Motors are  mounted on a support block that interfaces with the front and back support plate. To avoid ghosting, either the support assembly could be tilted or shims can be used inside individual filter holders.

The support assembly shown on the right panel of  Fig.~\ref{fig9} provides mounting for filter wheels, drive mechanisms and the following accessories.
\begin{itemize}

    \item [$\diamond$]	Four M4 jacking screws at the corner of the base for tip-tilt adjustment.
    \item [$\diamond$]	Nudger block adjustment for in-plane moment.
    \item [$\diamond$]	Temperature sensors at the base and at support block of the stepper motor.
    \item [$\diamond$]	Connectors for motor and other switches.
    \item [$\diamond$]	Switch assembly for home position and filter counting.
    \item [$\diamond$]	An opening at front and back support plate for exchanging filter unit.

\end{itemize}

\section{H2RG Detector System}
The H2RG detector subsystem for the SCALES imager is nearly identical to the one for the IFS channel. Both detectors are James Webb Space Telescope flight spares that are hardwired for 4-channel, slow-mode readout [\citenum{rau14}]. In order to use these detectors in the higher flux environment of ground-based astronomy, a new firmware will be deployed for reading full frames in 1 second or less. In some modes, there is a higher incident flux on the imager than on the IFS, which will require sub-framing and stripe-mode readouts.  Other aspects of the detector subsystems, including the mount design, thermal design and electronics, are identical between the imager and the IFS.

\section{Conclusions}
SCALES is a cryogenic instrument for doing a spatially resolved  high-resolution  imaging and spectroscopy of extrasolar planets and other celestial objects in thermal infrared. The foroptics of the instrument  utilises a standard coronagraph design that incorporates focal plane masks to block on-axis starlight and a downstream Lyot stops to prevent diffracted photons from reaching the detector. Imager channel is incorporated into SCALES baseline instrument to enhance its science capability, to take pupil images for alignment and to provide a timely replacement for the NIRC2 on Keck. It is optimized to cover a broad wavelength range 1-5 micron, though the detector has a high quantum efficiency down to 0.6 micron. The FoV of SCALES is $12.3''\times12.3''$ which is larger than the narrow field camera of NIRC2. The design is diffraction limited with 6~mas pixel sampling at 1 micron wavelength. A two-wheel filter mechanism equips the imager with standard YJHKLM filters and there is also a room to add more. A rotating pupil mask mechanism has been designed to cut down the  thermal infrared background and maximize the photon signal from M1 illuminated by the science target. The wavefront error is $< 60$~nm which is $3\times$ improvement over NIRC2. The maximum distortion within $10''\times10''$ FoV is $0.045''$ ($\sim 0.64\%$). Distortions are all low-order, which means it is easier to calibrate and correct them, either with a pinhole mask or star clusters. The aperture masking and coronagraphy are not included in the baseline design of the imager.

\acknowledgments 
 
Authors would like to acknowledge the funding support received from the Keck Observatory for the imaging channel design work. 

\bibliography{report} 
\bibliographystyle{spiebib} 

\end{document}